\documentclass[aps,prl,twocolumn,superscriptaddress,longbibliography]{revtex4-1}
\usepackage[T1]{fontenc}
\usepackage[latin9]{inputenc}
\usepackage{graphicx,amssymb,amsmath,xcolor}
\usepackage[unicode=true,
 bookmarks=false,
 breaklinks=false,pdfborder={0 0 0},pdfborderstyle={},backref=false,colorlinks=true]
 {hyperref}
\hypersetup{
 colorlinks}


\begin{document}

\title{\textit{Ab initio} calculations of erbium crystal field splittings in oxide hosts}

\author{Yogendra Limbu}
\affiliation{Department of Physics and Astronomy, University of Iowa, Iowa City, Iowa 52242, USA}
\author{Yueguang Shi}
\affiliation{Department of Physics and Astronomy, University of Iowa, Iowa City, Iowa 52242, USA}
\author{Joseph Sink}
\affiliation{Department of Physics and Astronomy, University of Iowa, Iowa City, Iowa 52242, USA}
\author{Tharnier O. Puel}
\affiliation{Department of Physics and Astronomy, University of Iowa, Iowa City, Iowa 52242, USA}
\author{Durga Paudyal}
\affiliation{Department of Physics and Astronomy, University of Iowa, Iowa City, Iowa 52242, USA}
\affiliation{Ames National Laboratory of the US DOE, Iowa State University, Ames, IA 50011, USA}
\affiliation{Department of Electrical and Computer Engineering, Iowa State University, Ames, IA 50011, USA}
\author{Michael E. Flatt\'e}
\affiliation{Department of Physics and Astronomy, University of Iowa, Iowa City, Iowa 52242, USA}
\affiliation{Department of Applied Physics, Eindhoven University of Technology, Eindhoven, The Netherlands}

\begin{abstract}
\noindent 
We present an effective \textit{ab initio} method to calculate the crystal field coefficients of an erbium (Er$^{3+}$) ion experiencing different local site symmetries in several wide-band-gap oxides, and then evaluate  crystal field splittings of these Er$^{3+}$ ions for their ground and excited states.
The optical transitions between the ground state ($Z$)  and excited state ($Y$) manifolds of the environmentally shielded $4f$ states of these Er$^{3+}$ ions have wavelengths $\sim$ 1.5 $\mu$m
and thus have
potential applications to quantum communications and quantum memories. These results are in excellent agreement with recent low-temperature measurements, provided the inadequate calculation of the $4f$ shell screening is adjusted by reducing the radial extent of the $4f$ wavefunctions by approximately a factor of 2.
\end{abstract}

\maketitle

Rare earth (RE) ions,  substituted in wide band gap insulating materials~\cite{grant2024Optical, thiel2011rare, kolesov2012optical}, can enhance the efficiency and speed of quantum information transmission over long distances and thus enable advances in quantum  information science (QIS)~\cite{stevenson2022erbium, zhong2019emerging,kunkel2015rare, Ji2024Nano,kenyon2002recent}. The crystal field splitting of localized 4$f$ electrons causes sharp optical, electric, and magnetic dipole
transitions~\cite{liu2005spectroscopic} favorable for QIS. 
The  $4f$ transitions of Er$^{3+}$ ions draw particular attention due to long optical and spin coherence times, controllable $4f$ linewidth broadening, and  emission in the telecommunications band~\cite{grant2024Optical, stevenson2022erbium,  thiel2011rare}. 
The Er$^{3+}$ ion, which  emits a  light  of $\sim$ 1.5 $\mu$~m from the \(4f\) - \(4f\) transition~\cite{galagan2006spectral}, has  diverse applications when hosted in wide band gap oxides, e.g., Y$_2$SiO$_5$ for quantum memory~\cite{thiel2012optical, lim2018coherent, jobez2015coherent} and Y$_2$O$_3$ for  spin-photon interfaces~\cite{bhandari2023distinguishing, mao2008synthesis}. When rare earth atoms are embedded into a host material, the lowered symmetry splits the high angular-momentum energy levels of the isolated atom into sets of multiplets~\cite{dresselhaus2007group}. 
For example, the ubiquitous Er$^{3+}$ ion, which has a vacuum ground state with $^{4}$I$_{15/2}$ character ($Z$ manifold) and excited state with $^{4}$I$_{13/2}$ character ($Y$ manifold),  often splits into eight ground-state (\(Z_1\) - \(Z_8\)) and seven excited-state doublets (\(Y_1\) - \(Y_7\)). 
As quantum-coherent phenomena often require controlled optical coupling between individual states, \textit{i.e.} a $Y_1 \rightarrow Z_1$ transition, knowledge of the crystal field splittings (CFS) informs decisions about bandwidth, cooling requirements, and performance for quantum operations.
Current experimental literature relies on solving an effective total Hamiltonian iteratively  to fit crystal field coefficients (CFCs)~\cite{10.1063/1.455853, duan2007vacuum}. There is a critical need to accurately predict CFCs and CFSs of RE ions hosted in wide-band-gap materials, preferably from \textit{ab initio} calculations with a minimal number of adjustable parameters. Development of a universal method is especially desirable so that inverse design techniques can identify potential QIS materials from  predicted CFCs and crystal field splittings.

Here we report the development of an effective method of calculating CFCs of Er doped into wide band gap oxides from {\it ab initio} (density functional theory, DFT)  calculations. Comparison with recent low-temperature  measurements~\cite{stevenson2022erbium} of CFSs of Er in MgO, TiO$_2$, ZnO, CaWO$_4$, and PbWO$_4$ show excellent agreement if the radial extent of the $4f$ wavefunction is modified by a single factor $\sim$ 2, which is nearly identical across these materials, and represents the limitation of DFT in calculating the radius of the strongly-correlated, partially-filled $4f$ shell of Er. 

Calculations of CFCs from DFT~\cite{novak2013crystal,novak2013, mollabashi2020crystal, ning2007density} have been attempted before, however  complexities arise from \(4f\)-\(4f\) self interaction and hybridization of $4f$ states with other states. This problem can partially be resolved by placing the $4f$ electrons in the core, forcing their electron density to have spherical symmetry~\cite{novak2013crystal,novak2013, mollabashi2020crystal}.  
Some\cite{novak2013crystal,novak2013,mollabashi2020crystal} have introduced
an empirical parameter to remove the hybridization between $4f$ states and other states using a Wannier function analysis. The empirical parameter is very different for different RE ions 
and its determination requires delicate analysis of the hybridization emerging from the calculation. 
An alternative method relies on an effective core potential using Gaussian basis sets~\cite{ning2007density}, however the $4f$ tail may still  hybridize unphysically with other valence states,  requiring an additional density-dependent pseudopotential term.  Here we calculate the CFCs within the $4f$ core approximation using a single physically-justifiable adjustable parameter, $\epsilon_{\omega, q}$, the dielectric constant. 

Once these CFCs are obtained the lanthanide ion energy spectrum was calculated using the {\it qlanth} code \cite{qlanth}, which includes terms representing the following interactions and relativistic corrections: spin-orbit, electrostatic repulsion, spin-spin, crystal field and spin-other-orbit as described in \cite{10.1063/1.455853}. {\it qlanth} now has major improvements over previous versions ({\it ie.} Ref. \cite{PhysRevB.86.125102}); it has fixed errors found in the term that includes configuration interaction via the Casimir operator of SO(3), has implemented the non-orthogonal terms for the three-body effective operators as discussed in \cite{Judd_Crosswhite_84}, has corrected some typos as well as improved accuracy in the Marvin integral terms, and implemented significant optimizations in the code itself. With these changes {\it qlanth}  reproduces the CFSs of Ref. \cite{10.1063/1.455853} with an agreement ($\lesssim30\text{ cm}^{-1}$ throughout all lanthanides).

\begin{figure}[!ht]
\includegraphics[width=0.99\columnwidth]{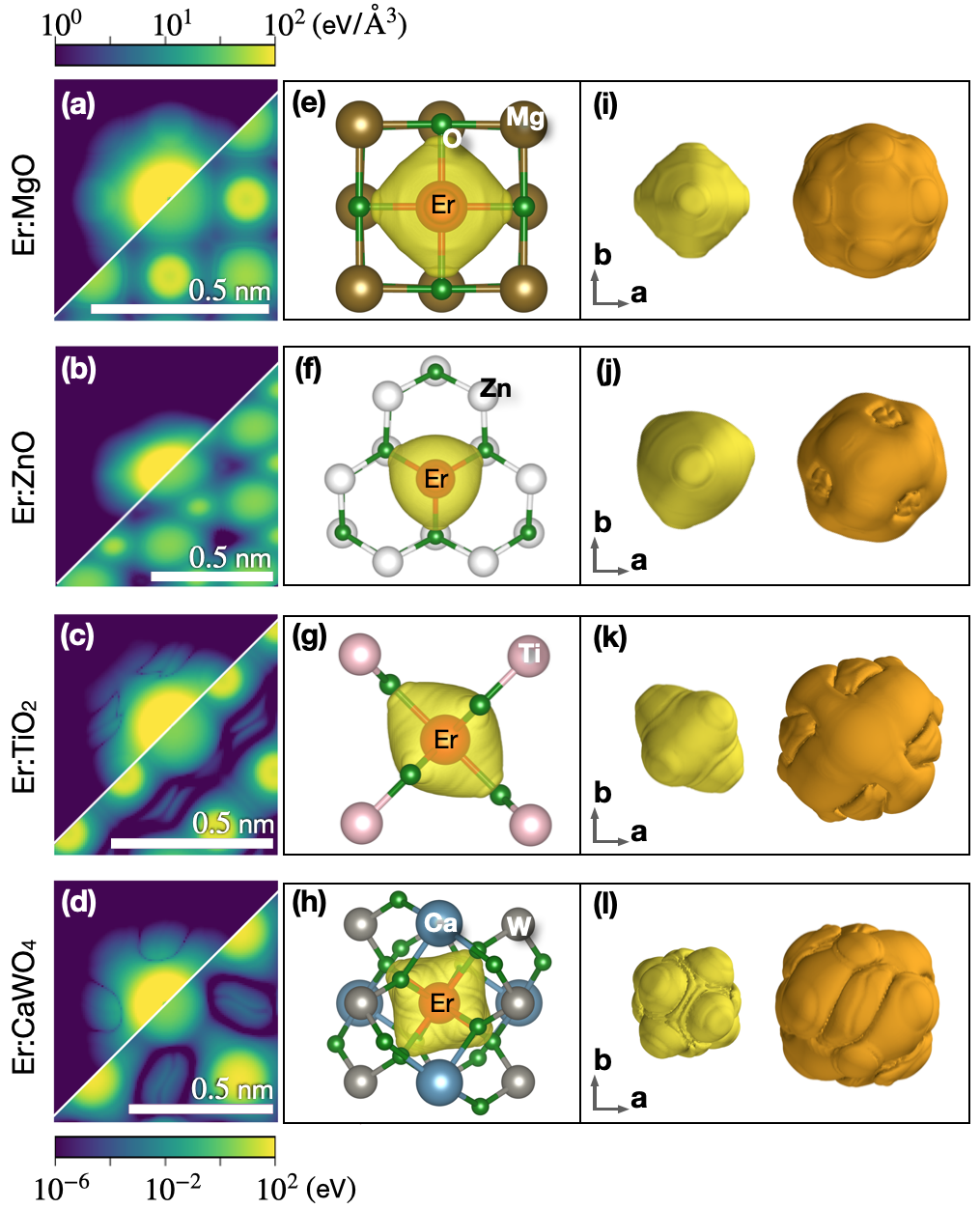}
\caption{(a)-(d) Two-dimensional views of the crystal field potential $V_\text{CF}(\textbf{r})$ and the same potential multiplied by the $4f$-electrons probability density of Er (upper diagonal, in $\text{eV}/$\AA$^3$).  (e)-(h) show the crystal structure surrounding the doped Er ion and the probability density of the 4$f$-electrons at  1.58 $\text{ eV}$/\AA$^3$ (yellow isocontour); (i)-(l) illustrate the probability density by plotting it at $10^{-2} \text{ eV}$/\AA$^3$ (left) and $10^{-6} \text{eV}$/\AA$^3$ (right) isocontours, omitting the depiction of the crystal.}
\label{figure1}
\end{figure}

The effective Hamiltonian of a single RE atom in the crystal environment is ~\cite{novak2013crystal,novak2013}: \(H_{\text{eff}}\) = \(H_{FI}\) + \(H_{CF}\), where  $H_{FI}$ and $H_{CF}$ are free ion and one-particle crystal field Hamiltonians. The free ion  Hamiltonian  of 4$f^N$ configuration, including two and three body operators in the configuration interactions is~\cite{10.1063/1.455853}:
\(H_{FI} = H_0 + \sum_{k}^6 F^kf_k + \zeta_{4f}A_{SO} + \alpha L(L+1) +  \beta G(G_2) + \gamma G(G_7) + \sum_{i} t_iT^i+ \sum_{h} m_hM^h + \sum_{f} p_fP^f\),
where $H_0$ is  spherically symmetric one-electron Hamiltonian, $F^k$ and $f_k$ ($k$ = 0, 2, 4, 6) are electrostatic integral and angular part of the electrostatic interaction in two-body configuration, $\zeta_{4f}$ and $A_{SO}$ are spin-orbit integral and angular part of spin-orbit interaction, $\alpha$, $\beta$, and $\gamma$ are the Trees parameters (correction terms, which are obtained from \textit{ab initio} method or fitting experimental data~\cite{10.1063/1.455853}) associated with the two-body interaction operators~\cite{rajnak1963configuration, trees1951configuration}, and $G(G_2$) and $G(G_7$) are the eigenvalues of Casimir's operators for the groups $G_2$ and $G_7$, whereas $L$ is  total orbital angular momentum~\cite{rajnak1963configuration,wybourne1965spectroscopic}. For the three  or more  $4f$ electrons, the three-particle configuration interaction is included by adding  $t_iT^i$ ($i$ = 2, 3, 4, 6, 7, 8) in two particles Hamiltonian~\cite{judd1966three, rajnak1963configuration}, where $T^i$ are the adjustable
parameters, which are obtained from  \textit{ab initio} method or fitting experimental
data~\cite{10.1063/1.455853,carnall1978energy}, and $t_i$ are three-particle operators. These atomic parameters are extracted from Ref.~\cite{10.1063/1.455853}. Further, $M_h$ ($h$ = 0, 2, 4)  contains  spin-spin and spin-other-orbit relativistic corrections, known as Marvin integrals~\cite{marvin1947mutual}, while $P^f$ ($f$ = 2, 4, 6) represents the electrostatically correlated spin-orbit perturbation of  two-body magnetic corrections.  $m_h$ and $p^f$ are operators associated with these magnetically correlated corrections.

The crystal field   potential~\cite{wybourne1965spectroscopic} is: \(V_{CF} = \sum_{k, q} B_{q}^k (r) C_{q}^k (\theta, \phi)\),
where $B_q^k$(r) and $C_{q}^k$($\theta$, $\phi$) are the radial and spherical components of $V_{CF}$.   The position independent $B_q^k$ coefficients,  which depend on the local site symmetry of RE ion in the crystal~\cite{10.1063/1.455853,ning2007density} are: 
\begin{equation}
    B_q^k = \frac{2k+1}{4\pi}\int { V_{CF}(r, \theta, \phi) R^{2}_{4f}(r) C^{k*}_{q}(\theta, \phi) r^{2}dr \sin{\theta}} d{\theta} d{\phi}, 
\nonumber
\end{equation}
where $R^{2}_{4f}(r)$ and $C^{k*}_q (\theta, \phi)$ indicate the square of the $4f$ radial wave function  and the complex conjugate of the spherical tensor. The $4f$ radial wave function of the RE ion, which depends on the dielectric constant of the material and position of $4f$ electrons is: \(R_{4f} (r,\epsilon_{\omega,\mathbf{q}\sim a_0}) = A r^3 e^{-rZ_{\text{eff}}/{na_0\epsilon_{\omega,q}}}\),
where  $A$, $n$, $a_0$, and $Z_{\text{eff}}$  are the amplitude of $4f$ wave function, principal quantum number, the Bohr radius, and effective nuclear charge of $4f$ orbital, respectively, in the RE ion. The orbital exponent of $4f$ electron ($\zeta_{n, l, m}$) of Er, which is related to the screening constant, is taken from Ref.~\cite{clementi1967atomic} for calculating the effective nuclear charge of $4f$ orbital  using relation $Z_{\text{eff}}$ = $n\zeta_{n, l, m}$, yielding a value of 27.9784. Here the dielectric constant, $\epsilon_{\omega, q}$, which depends upon the frequency and crystal structure, and controls the spreading of the $4f$ radial wave function, is carefully optimized by fitting the calculated and the experimental energy levels. The optimized $\epsilon_{\omega, q}$ for MgO, ZnO, TiO$_2$,  PbWO$_4$, and CaWO$_4$ are 1.96, 1.90, 2.00,  2.24, and 2.24, respectively. The corresponding $\epsilon_{\omega, q}$ are then used to extract CFCs from the self-consistent DFT charge densities and local potentials.

In addition, we performed  maximally localized Wannier functions  (Wannier90~\cite{mostofi2008wannier90}) calculations and extracted maximally-localised Wannier $4f$ functions with $4f$ electrons as valence electrons. These extracted $4f$ Wannier functions are then compared with hydrogenic $4f$ radial wave functions for different values of $\epsilon_{\omega,q}$. The extracted  Wannier function is in good agreement with the hydrogenic wave function when epsilon is in between 3 and 4. This is also confirmed by the ionization energy of Er$^{3+}$ ($\sim$ 42.42 eV~\cite{johnson2017lanthanide, lang2010ionization}), which is expressed as, \(E_{i}={m_e e^4Z_{\text{eff}}^2}/{(4\pi\epsilon_{\text{eff}})^2{\hbar^2}}n^2\), where  $m_e$ and  $\epsilon_{\text{eff}}$(= $\epsilon_{\omega, q}\epsilon_0$) are the electron mass and permittivity. The  resulting dielectric constant, $\epsilon_{\omega, q}$, is 3.96 in agreement with the Wannier90 value (Fig.~3~\cite{limbuprb}).

\begin{table}[ht!]
\hfill{}
\begin{tabular}{l c c c c c}
\toprule
$B_q^k$ & MgO   & ZnO  & TiO$_2$  & CaWO$_4$  & PbWO$_4$ \\
\hline
$B_{0}^2$ &    &  -260.96  & 74.19 & 846.34  & 872.91\\
$B_{\pm{2}}^2$ &  & & $\pm i$474.97 &  &  \\
$B_{0}^4$ & 660.96 & 199.04 & -493.49 & 187.43  & 90.08\\
$B_{\pm{2}}^4$ &   &   & $\pm i$146.80 &  &   \\
$B_{\pm{3}}^4$ &   &  -$i$92.07 & &  &   \\
$B_{\pm{4}}^4$ &  395.00  &  & 310.25 & 
$\left(
\begin{array}{c}
     192.40\\
     \mp i56.58
\end{array}
\right)$  & 
$\left ( 
\begin{array}{c}
114.55\\
\pm i2.79
\end{array}
\right )$
\\
\(B_{0}^6\) & -78.57  &  -240.17 & 466.69 &  87.40 & 138.50\\
$B_{\pm{2}}^6$ &   &  & $\pm i$85.65 &  &  \\
$B_{\pm{3}}^6$ &   & -$i$228.92 &  &  &  \\
$B_{\pm{4}}^6$ & 146.98 &  & 106.82 & 
$\left ( \begin{array}{c}
     311.06\\
     \mp i424.36
\end{array} \right)$   
& $\left ( \begin{array}{c}
     182.07\\
     \mp i352.72
\end{array}\right)$\\
$B_{\pm{6}}^6$ &  & 233.60 & $\pm i$254.93 &   & \\
\hline
\hline
\end{tabular}
\hfill{}
\caption{extracted CFCs (units of cm$^{-1}$) of Er doped in wide band gap oxides with variable local site symmetry, e.g., MgO with $O_h$, 
ZnO with $C_{3v}$, TiO$_2$ with $D_{2h}$, and CaWO$_4$ and PbWO$_4$ with $S_4$, from DFT calculations.}
\label{table1}	
\end{table}

The DFT calculations are performed   using the Vienna \textit{Ab-initio} Simulation Package~\cite{hafner2008ab,kresse1996efficiency}. Augmented plane wave pseudopotentials~\cite{blochl1994projector} and Perdew-Burke-Ernzerhof  functionals~\cite{ernzerhof1999assessment}  are used. A sufficient plane wave cut off energy (500 eV) and $\Gamma$ centered \textit{k}-mesh (maximum of 10 $\times$ 10 $\times$ 10) are used for the Brillouin zone sampling.
The $4f$ electrons are frozen in the core and  the $6s^2$, $5p^6$, and $5d^1$ electrons are considered in the valance shell.  From non-spin polarized calculations the self-consistent  charge densities and local potentials were generated to produce the $B_q^k$ considering the local potentials as  crystal fields.

In order to study the crystal field splitting of 4\(f\) states of Er\(^{3+}\) ions, we focused on the oxides MgO, ZnO, TiO$_2$, CaWO$_4$ and PbWO$_4$, with variable point group symmetry. MgO, ZnO, and TiO$_2$ have cubic, hexagonal, and rutile (tetragonal) crystal structures with $O_h$, $C_{6v}$, and $D_{4h}$ point groups and $Fm{\overline{3}m}$, P6$_3$mc, and P4$_2$/{mnm} space groups, respectively, whereas both CaWO$_4$ and PbWO$_4$ have tetragonal crystal structure with $C_{4h}$ point and I4$_1$/a space groups. 
The  optimized lattice parameters for MgO ($a$ = 4.24~\AA), ZnO ($a$ = 3.28 \AA~and $c$ = 5.30 \AA),  TiO$_2$ ($a$ = 4.66 \AA~and $c$ = 2.96 \AA), CaWO$_4$ ($a$ = 5.29 \AA~and $c$ = 11.43 \AA), and PbWO$_4$ ($a$ = 5.51 \AA~and $c$ = 12.13 \AA) are in good agreement with the available literature~\cite{richter2013concentration,meyer2003density,cavalcante2012electronic,zhang1998electronic,korner2011density}.

Calculated band gaps for TiO$_2$, CaWO$_4$, and PbWO$_4$ from a standard hybrid functional (HSE06) are 3.04, 5.39, and 4.30 eV (Fig.~4~\cite{limbuprb}), which are in good agreement with the corresponding experiments~\cite{mikhailik2004one, lacomba2008optical,serpone2006band}. However, the  standard hybrid functional was unable to reproduce the experimental band gaps of MgO and ZnO. In these cases, the hybrid
functional calculations with Hartree Fock mixing parameters of 0.45 for MgO and 0.38 for ZnO show band gaps of 7.65 eV and 3.31 eV (Fig.~4~\cite{limbuprb}), which are indeed in good agreement with corresponding experiment~\cite{whited1973exciton,srikant1998optical}.
These wide band gap non magnetic host oxides with nuclear spin-free isotopes ~\cite{stevenson2022erbium} become excellent hosts for Er$^{3+}$ ions to provide sharp 4$f$ - 4$f$ transitions, exhibiting long spin and optical coherence times needed for quantum storage and spin-photon interface.

\begin{figure}[t]
\includegraphics[width=0.99\columnwidth]{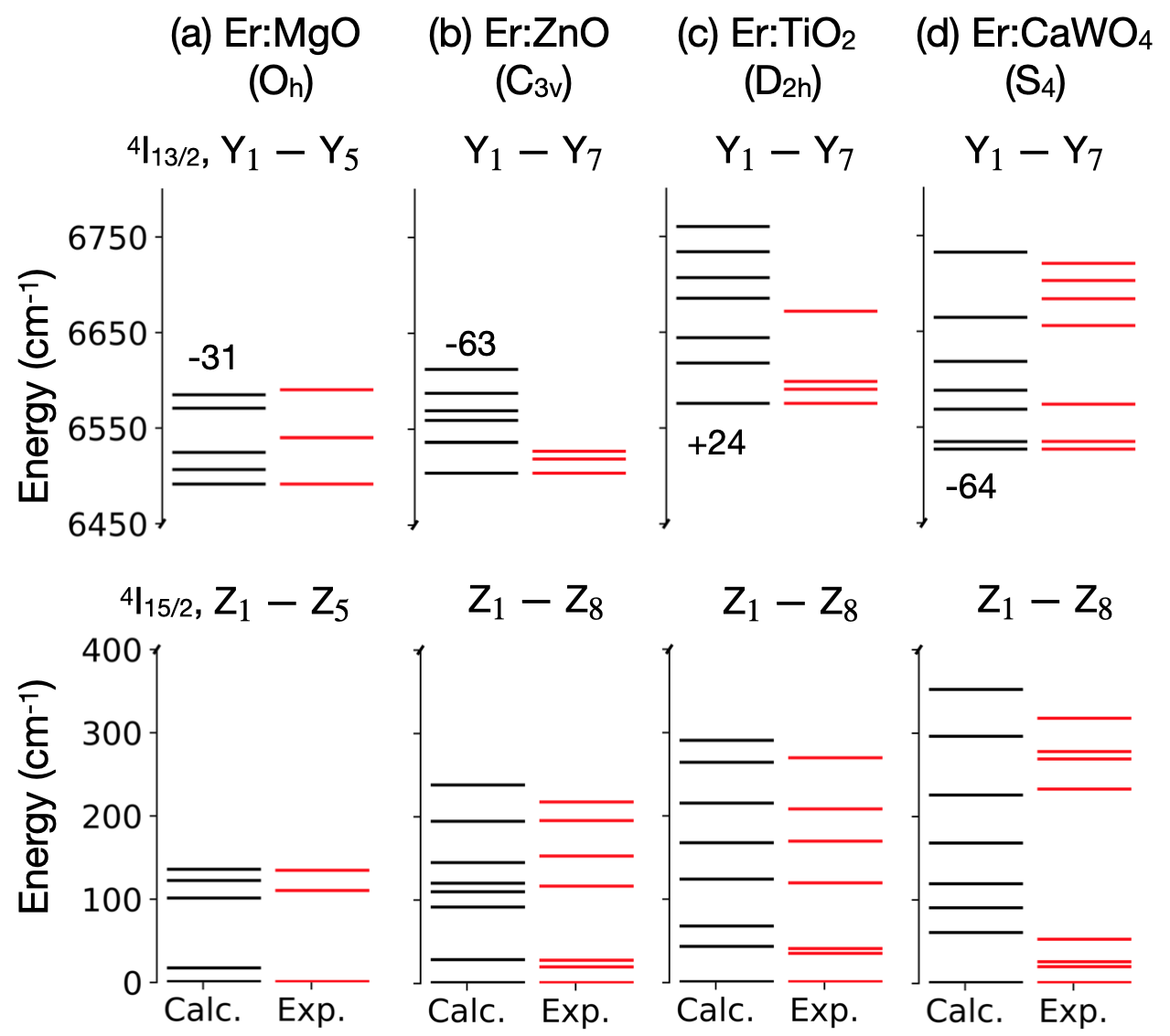}
\caption{The ground ($^4 \text{I}_{15/2}$) and first excited ($^4 \text{I}_{13/2}$) multiplets splittings of the 4$f$ states of Er$^{3+}$ ion in (a) MgO, (b) ZnO, (c) TiO$_2$, and (d) CaWO$_4$. 
The eight Kramers-pairs levels from the ground state, $\text{Z}_1$ - $\text{Z}_8$, and seven levels from the first excited state of Er$^{3+}$,
$\text{Y}_1$ - $\text{Y}_7$, are identified in all cases, except for MgO, in which the multiplet is split into five energy levels. 
In the $^4 \text{I}_{13/2}$ splittings, we have adjusted the calculated values to aligned with the $\text{Y}_1$ experimental values (e.g., -31 cm$^{-1}$ for Er:MgO), as the gap between the multiplets is not a property from the crystal field and can be adjusted with the Er ion parameters.
}
\label{figure5}
\end{figure}

Now we describe CFCs calculations and results using Wybourne notation and cm$^{-1}$ units~\cite{wybourne1965spectroscopic}. In MgO, the Er  site has $O_h$ local site symmetry \cite{borg1970spin,Baker_1976}. From  group theory, only  the $B_q^k$  for  $k$  = 4 and 6, and $q$ = 0 and $\pm$4  are non-zero, satisfying the  relations {$B_{\pm{4}}^4$} = $\sqrt{\frac{5}{14}}B_{0}^4$ and  {$B_{\pm{4}}^6$} = -$\sqrt{\frac{7}{2}}{B_{0}^6}$~\cite{ning2007density}. All CFCs are positive, except for $B_0^6$.  In CaWO$_4$ and PbWO$_4$, the Er doped site favors $S_4$  local site symmetry~\cite{enrique1971optical}. Five non-zero CFCs: $B_{0}^2$,  $B_{0}^4$, $B_{\pm{4}}^4$, $B_{0}^6$, and $B_{\pm{4}}^6$ are identified in both materials as also suggested in   Refs.~\cite{wortman1971optical, mollabashi2020crystal}. The $B_{\pm{4}}^4$ and $B_{\pm{4}}^6$ have  real and imaginary components, while $B_0^2$, $B_0^4$, and $B_0^6$ have only the real components. 
The real components of CFCs follow the relation  Re[$B_q^k$] = Re[$B_{-q}^k$], whereas the imaginary components follow the relation, Im[$B_{-q}^k$] = -Im[$B_{q}^k$], except for $B_{\pm{4}}^4$ in PbWO$_4$ with Im[$B_{-q}^k$] = Im[$B_{q}^k$]. 
In ZnO, Er doped site favors $C_{3v}$ local site symmetry~\cite{stevenson2022erbium}. Six non-zero CFCs:  $B_0^2$, $B_0^4$, $B_{\pm{3}}^4$, $B_{0}^6$,  $B_{\pm{3}}^6$, and $B_{\pm{6}}^6$ are deduced as described in Refs.~\cite{divivs2008crystal, wybourne1965spectroscopic}. The coefficients:  $B_{\pm{3}}^4$ and $B_{\pm{3}}^6$ have only the imaginary components with the relation Im[$B_{q}^k$] = Im[$B_{-q}^k$]. In TiO$_2$, Er doped site has $D_{2h}$ local site symmetry~\cite{hamed2023dominant}. In this case, there are five real and four imaginary CFCs as also mentioned in Ref.~\cite{furrer1988neutron}. The real CFCs are $B_0^2$, $B_0^4$, $B_{\pm{4}}^4$, $B_0^6$, and  $B_{\pm{4}}^6$ and the imaginary CFCs are $B_{\pm{2}}^2$, $B_{\pm{2}}^4$, $B_{\pm{2}}^6$, and $B_{\pm{6}}^6$. The real and imaginary components follow the relation Re[$B_{q}^k$] = Re[$B_{-q}^k$] and Im[$B_{q}^k$] = -Im[$B_{-q}^k$]. Thus, in $S_4$, $D_{2h}$, and $C_{3v}$ local site symmetries, the 4$f$ splitting of  Er$^{3+}$ ions are due to the second, forth, and sixth order CFCs, whereas in $O_h$ local site symmetry, the 4$f$ splitting of  Er$^{3+}$ ion is due to the forth and sixth order CFCs.

The 4$f$ states of Er$^{3+}$ with O$_h$ local site symmetry splits the ground state into three quartets $\Gamma_8$ (0.00, 17.37, and 122.52 cm$^{-1}$)  and two doublets $\Gamma_7$ (101.32 cm$^{-1}$) and $\Gamma_6$ (135.95 cm$^{-1}$) (Fig.~6~\cite{limbuprb}) as also suggested in Refs.~\cite{bleaney1959new, borg1970spin} and observed experimentally ~\cite{stevenson2022erbium} (Fig.~\ref{figure5}(a)). We note that the order and position of doublets and quartets are sensitive to the ratio of fourth and six order CFCs ~\cite{white1962energy, chen1984superposition}.  We followed the irreducible representation mentioned in Ref.~\cite{bradley1976p}. Similarly, two  quartets (6537.66  and 6615.81 cm$^{-1}$) and three doublets (6522.43 cm$^{-1}$ ($\Gamma_6$), and 6555.57  and 6601.82 cm$^{-1}$ ($\Gamma_7$)) are identified in the first excited state.  The 4$f$ - 4$f$ transition from $Y_1$ to $Z_1$  is 6522.43  cm$^{-1}$. Without crystal field environment, the calculated 4$f$ - 4$f$ transition from the first excited state to the ground state is 6488.00 cm$^{-1}$ (6478.00 cm$^{-1}$ in the experiment~\cite{enrique1971optical}), which is 34.43 cm$^{-1}$ shorter than that of with the crystal field environment. This indeed shows the increase of wave length when the crystal environment is neglected.

In Er doped ZnO with $C_{3v}$ symmetry, the 4$f$ states of Er$^{3+}$ ion splits  into three $\Gamma_{5, 6}$ and five $\Gamma_4$ Kramers doublets in the ground state (Fig.~6~\cite{limbuprb}), which are in good   agreement with the experiment (Fig.~\ref{figure5}(b)). Here, $\Gamma_5$ and $\Gamma_6$  form a doublet, while $\Gamma_4$ forms a separate doublet. In the ground state,  the  energy levels $Z_1$, $Z_4$, and $Z_7$ correspond to $\Gamma_{4,5}$, and the remaining energy levels $Z_2$, $Z_3$, $Z_5$, $Z_6$, and $Z_8$ correspond  to $\Gamma_4$.
Similarly,  seven  energy levels,  2$\Gamma_{4,5}$ + 5$\Gamma_4$, are found in the first excited state (Fig.~\ref{figure5}(b)), where $Y_2$ and $Y_6$ are $\Gamma_{5,6}$, and the remaining $Y_1$, $Y_3$, $Y_5$, and $Y_7$ are $\Gamma_4$. In the experiment, only  seven and three multiplets  are visualized in the ground and  first excited states ~\cite{stevenson2022erbium}. The calculated 4$f$ - 4$f$ transition from $Y_1$ to  $Z_1$ is 
 6567.52 cm$^{-1}$, which is 63.32 cm$^{-1}$ larger than  the  experiment~\cite{stevenson2022erbium}.

Er doping in TiO$_2$ has $D_{2h}$ local site symmetry. The 4$f$ states form eight and seven Kramers doublets as represented by $\Gamma_5$ (Fig.~6~\cite{limbuprb}) as mentioned in Ref.~\cite{goodman1991crystal} in the ground  and first excited states, which are in fair agreement with the experiment~\cite{stevenson2022erbium, phenicie2019narrow}(Fig.~\ref{figure5}(c)). In the experiment, only seven and four multiplets  are observed  in the ground and first excited states~\cite{stevenson2022erbium, phenicie2019narrow}. The 4$f$ - 4$f$ transition from  $Y_1$  to $Z_1$  is  6552.43 cm$^{-1}$, which is (23.27 cm$^{-1}$) smaller than the experiment~\cite{stevenson2022erbium}. The $D_{2h}$ local site symmetry does not allow a permanent electric dipole
moment~\cite{phenicie2019narrow,stevenson2022erbium}, revealing the 4$f$ - 4$f$ transition is mainly due to magnetic dipole transition. 

In the case of Er doped CaWO$_4$ with $S_4$ local site symmetry, the 4$f$ states also form eight  (4$\Gamma_{5,6}$ + 4$\Gamma_{7,8}$) and seven  (4$\Gamma_{5,6}$ + 3$\Gamma_{7,8}$) Kramers doublets in the ground and first excited states (Fig.~6~\cite{limbuprb}), which are also in fair agreement with the experiment~\cite{wortman1971optical} (Fig.~\ref{figure5}(d)). In the ground state, the energy levels $Z_1$, $Z_3$, $Z_6$, and $Z_7$ are  $\Gamma_{5,6}$, while the other four states $Z_2$, $Z_4$, $Z_5$, and $Z_8$ are $\Gamma_{7,8}$. The energy levels $Y_1$, $Y_3$,  $Y_6$, and  $Y_7$ correspond to $\Gamma_{5,6}$, while the other three states $Y_2$, $Y_4$, and $Y_5$ are $\Gamma_{7,8}$ in the first excited state. The  4$f$ - 4$f$ transition from $Y_1$ to $Z_1$  is  6591.30 cm$^{-1}$, which is 64.30 cm$^{-1}$ larger than the experiment~\cite{wortman1971optical} (Fig.~\ref{figure5}(d)). PbWO$_4$ has the same crystal structure as CaWO$_4$. Thus, eight  and  seven Kramers doublets  are identified in the ground and first excited states. The  4$f$ - 4$f$ transition from $Y_1$ to $Z_1$ is  6570.05 cm$^{-1}$, which shows a good agreement with experiment~\cite{stevenson2022erbium}.

In conclusion, we have developed an effective \textit{ab initio} method of calculating crystal field coefficients and energy splitting of 4\textit{f} states of Er$^{3+}$ ions in wide band gap oxides with different local site symmetries, e.g., $O_h$, $C_{3v}$, $D_{2h}$, and $S_4$. The  calculated band gaps of representative oxides with different point group symmetry, such as MgO with $O_h$, ZnO with $C_{6v}$, CaWO$_4$ and PbWO$_4$ with $C_{4h}$, and TiO$_2$ with $D_{4h}$, using density functional theory (DFT) incorporating a fraction of exact exchange, are in good agreement with available experimental values. The crystal field coefficients of Er$^{3+}$ ions in these oxides are extracted from self-consistent charge densities of non spin polarized DFT calculations within the core approximation. These coefficients are then fed into an effective Hamiltonian to generate the crystal field splitting of Er$^{3+}$. The resulting crystal field multiplets are in good agreement with available experimental values. Further details of the calculations presented here are available in a simultaneously submitted publication\cite{YogendraCrystalPRB}. 

\begin{acknowledgments}
This work is supported by the U.S. Department of Energy, Office of Science, Office of Basic Energy Sciences under Award Number DE-SC0023393.
\end{acknowledgments}

\end{document}